\documentclass[twocolumn,times]{aastex62}

\usepackage{graphicx}
\usepackage{amsmath}

\usepackage{longtable} 
\usepackage{natbib} 

\begin{document}

\title{\textcolor{black}{An Accretion}-Modulated Internal Shock Model for Long GRBs}

\author[0000-0002-2516-5894]{R.~Moradi}
\affiliation{State Key Laboratory of Particle Astrophysics, Institute of High Energy Physics, Chinese Academy of Sciences, Beijing 100049, China.}
\email{rmoradi@ihep.ac.cn}

\author{C.~W.~Wang}
\affiliation{State Key Laboratory of Particle Astrophysics, Institute of High Energy Physics, Chinese Academy of Sciences, Beijing 100049, China.}
\affiliation{University of Chinese Academy of Sciences, Chinese Academy of Sciences, Beijing 100049, China}

\email{cwwang@ihep.ac.cn}

\author{ E. S. Yorgancioglu}
\affiliation{State Key Laboratory of Particle Astrophysics, Institute of High Energy Physics, Chinese Academy of Sciences, Beijing 100049, China.}
\affil{University of Chinese Academy of Sciences, Chinese Academy of Sciences, Beijing 100049, China}

\author{S. N. Zhang}
\affiliation{State Key Laboratory of Particle Astrophysics, Institute of High Energy Physics, Chinese Academy of Sciences, Beijing 100049, China.}
\affiliation{University of Chinese Academy of Sciences, Chinese Academy of Sciences, Beijing 100049, China}

\begin{abstract}

We introduce the \textit{\textcolor{black}{Accretion}-Modulated Internal Shock model}
(\textcolor{black}{A}MIS) as a possible framework for explaining the observational
properties of long gamma-ray burst (GRB) prompt emission. In this scenario, the
envelope of the prompt light curve follows the
\textcolor{black}{time-dependent mass-supply history to the central engine,
associated with stellar collapse and, where applicable, fallback accretion},
whose \textcolor{black}{early-time onset can be approximated by}
$\dot{M}\propto t^{0-1/2}$ and which subsequently \textcolor{black}{may} decay as
$\dot{M}\propto t^{-5/3}$, producing a photon count rate with a single
fast-rise–exponential-decay (FRED)–like profile.
\textcolor{black}{In gereral, the prompt-emission envelope is regulated by a time-dependent
mass supply to the central engine, while internal shocks produce the rapid
variability.} Since we only aim to introduce this framework here, we focus on the simplest
single-FRED shape of the prompt emission profiles, while more complex cases
involving multiple episodes and interacting shocks will be explored in
forthcoming studies. The model indicates correlations between spectral
evolution, FRED-pulse narrowing at high energies, and the
\textcolor{black}{mass-supply--controlled} envelope. Stochastic Lorentz-factor
variations of ejected mass- or rate-driven shells, superimposed on the
\textcolor{black}{Accretion}-Modulated envelope, explain the coexistence of smooth
global trends and irregular short-timescale features, such as the widths of
individual pulses in long GRB light curves, offering diagnostic tools for
probing the inner engine activity.

\end{abstract}


\section{Introduction} \label{sec:intro}

The light curves of gamma-ray bursts (GRBs) in their prompt emission episode usually display rapid, multi-peaked variability, traditionally interpreted within the framework of internal shocks (IS) among the succession of ultra-relativistic shells emanated from the inner engine \citep{1994ApJ...430L..93R,1996AIPC..384..782S,1997MNRAS.287..110S,1997ApJ...490...92K,1998MNRAS.296..275D,zhang2018physics,2022Galax..10...38B,2009A&A...498..677B,2014A&A...568A..45B}. Thanks to the high resolution of missions such as the Burst and Transient Source Experiment (\textit{BATSE}; $20$–$2000$~keV) \citep{1985ICRC....3..343F}, Fermi’s Gamma-ray Burst Monitor (GBM; $8$~keV–$40$~MeV)\citep{2009ApJ...702..791M}, Konus-Wind ($20$–$20000$~keV)\citep{1995SSRv...71..265A}, the Hard X-ray Modulation Telescope (HXMT; $1$–$3000$~keV) \citep{HXMT_zhang_2018,HXMT_zhang_2020}, and the Gravitational wave high-energy Electromagnetic Counterpart All-sky Monitor \citep[GECAM; $6$–$6000$~keV;][]{GEC_INS_Li2022,GEC_INS_An2022,HEBS_INS_Zhang2023,GTM_INS_wang2024}, prompt-phase light curves now provide valuable information not only about the radiation mechanisms but also about the temporal behavior of \textcolor{black}{mass and energy injection associated with the central engine} \citep{1997ApJ...490...92K,2003MNRAS.342..587D,2012A&A...542L..29H,Maxham-Zhang09}. Empirically, several studies have reported systematic correlations among temporal observables such as pulse waiting times, full width at half-maximum (FWHM), and peak fluxes, in addition to stochastic variability. Such correlations suggest that the shell-production process is modulated by a varying driver, possibly linked to the structure or evolution of the inner engine, rather than purely random fluctuations \citep{2001ApJ...550..410M,2002ApJ...572L.139N,2012ApJ...744..141B,2015ApJ...801...57G,2026JHEAp..4900456M}.

Internal shock models successfully capture the observed nonthermal spectra and
variable pulse structure \citep[e.g.,][]{1997ApJ...490...92K,
2002ApJ...572L.139N,2014A&A...568A..45B}. However, the systematic temporal trends
reported in many bursts are still not well accounted for within IS models
\citep[see e.g.,][]{2024ApJ...977..155M}. At the same time, core-collapse
simulations show that fallback of bound stellar debris onto the compact remnant
produces a time-dependent mass-supply rate \textcolor{black}{whose late-time
evolution follows} $\dot M(t)\propto t^{-5/3}$ and \textcolor{black}{may exhibit an
early-time onset consistent with} $\dot M(t)\propto t^{0-1/2}$
\citep{1989ApJ...346..847C,2001ApJ...550..410M,2008MNRAS.388.1729K,
2024ApJ...961..212J}. This behavior sustains long-duration energy injection, and
\textcolor{black}{when coupled to a relativistic outflow whose luminosity follows
the time-dependent mass supply to the central engine,} produces a
fast-rise–slow-decay (FRED)–like envelope. Therefore, it can imprint broad
temporal trends on the prompt light curve. Stochastic variations in the Lorentz
factor of the ejected shells riding on top of this
\textcolor{black}{mass-supply–regulated} envelope can then produce the observed
fine-scale multi-peaked structure. In this framework, namely the
\textcolor{black}{Accretion}-Modulated internal shock (\textcolor{black}{A}MIS)
model, pulse properties such as FWHM, waiting times, and amplitude evolution
emerge as the combined result of \textcolor{black}{time-dependent mass-supply
histories (fallback-dominated where applicable)} and internal shock physics.

The fallback-accretion--driven framework has been utilized to account for the
early afterglow flares in some GRBs \citep[e.g.,][]{2013ApJ...767L..36W,
2013ApJ...779...28G}. \textcolor{black}{Signatures consistent with fallback-regulated
mass-supply decay have also been reported in the afterglow of ultra-long GRBs
such as GRB~250702B \citep{2025arXiv250926283Z}}. However, direct application to
the prompt phase remains limited, and a model connecting
\textcolor{black}{mass-supply--controlled (fallback-dominated where applicable)}
broad envelopes with variable internal shocks has not yet been fully developed.
In this paper, we introduce the \textit{\textcolor{black}{Accretion}-Modulated
Internal Shock model} as a general framework for long GRB prompt emission,
\textcolor{black}{in which a phenomenological, time-dependent mass-supply history
to the central engine---arising from stellar collapse and including fallback as
a physically motivated limiting case---is mapped onto shell ejection and
internal-shock dissipation.} We discuss its predictions for temporal and
spectral observables, and highlight how it can account for both global
light-curve trends and fine-scale stochastic variability.

The paper is organized as follows. In Section~\ref{sec:fallbackjetmodel}, we review
fallback accretion in collapsars and its role in \textcolor{black}{regulating the
time-dependent mass supply that shapes} the broad temporal envelope of GRB
prompt emission. In Section~\ref{sec:FMIS} we introduce the mathematical
foundation of the \textcolor{black}{A}MIS model by describing shell-collision
dynamics. We further present in Section~\ref{sec:FMIS1} the framework that maps
the \textcolor{black}{time-dependent mass-supply history} onto shell properties
within two representative modulation scenarios. We then examine the spectral
evolution of pulses and derive analytic estimates for band-limited pulse widths
in Section~\ref{sec:energy-width}. Section~\ref{sec:conclusions} summarizes our
findings and discusses implications for central-engine physics and progenitor
structure, together with caveats and possible avenues for future improvement.

\textcolor{black}{\section{Fallback-Regulated Mass Supply and Jet Luminosity in Collapsars}\label{sec:fallbackjetmodel}}

Long GRBs are widely interpreted as the collapse of massive, rapidly rotating
stars that form either a black hole or a neutron star, capable of launching
relativistic jets. In the collapsar framework, two evolutionary channels are
proposed: Type~I and Type~II collapsars
\citep[e.g.,][]{1999ApJ...524..262M,1999ApJ...518..356P,1999ApJ...526..152F,2007ApJ...670.1247W}.
\textcolor{black}{
The mass supply to the central engine is intrinsically
time-dependent and may involve both prompt collapse-driven mass accretion and
delayed fallback of bound stellar material, depending on the progenitor structure
and explosion outcome \citep[e.g.,][]{2001ApJ...550..410M,2001ApJ...557..949N}.
}

In Type~I collapsars, the black hole promptly forms with a hyper-accreting disk of
typical accretion rates
\(\dot{M} \sim 10^{-2} - 1~M_\odot~\mathrm{s^{-1}}\),
depending on the progenitor structure, angular momentum, and disk viscosity
\citep{1999ApJ...524..262M,1999ApJ...518..356P}. Such high rates
\textcolor{black}{are capable of} efficiently powering relativistic jets via
neutrino annihilation or magnetohydrodynamic (MHD) processes. Assuming a
radiative efficiency
\(\epsilon_{\rm rad} \sim 0.01 - 0.2\), the isotropic-equivalent prompt
luminosities \textcolor{black}{can therefore} reach
\(L_{\rm prompt} \sim 10^{53}~\mathrm{erg~s^{-1}},\)
consistent with the majority of observed long GRBs. Immediate black hole
formation \textcolor{black}{is thus typically associated with} bright prompt
emission, while \textcolor{black}{any precursor activity is expected to be weak or
absent}, with additional variability arising primarily from jet propagation
effects.

Type~II collapsars occur when the initial explosion is weak or fails, leaving a
proto-neutron star while a significant fraction of the stellar envelope remains
gravitationally bound. \textcolor{black}{Subsequent fallback of marginally bound
material} then occurs on timescales
\(t_{\rm fb} \sim 1 - 10^3~\mathrm{s}\)
\citep{1989ApJ...346..847C,1999ApJ...526..152F,2008ApJ...679..639Z}. If sufficient
mass is accreted, the proto-neutron star collapses into a black hole,
\textcolor{black}{potentially enabling} the delayed launching or re-energization of
a relativistic jet. Typical fallback rates are smaller,
\(\dot{M} \sim 10^{-3} - 10^{-2}~M_\odot~\mathrm{s^{-1}}\),
for which neutrino cooling and annihilation are inefficient
\citep{1999ApJ...518..356P}, leading to lower radiative efficiency.
\textcolor{black}{The resulting prompt luminosities are therefore expected to lie in}
\[
L_{\rm prompt} \sim 10^{47} - 10^{49}~\mathrm{erg~s^{-1}},
\]
unless magnetic energy extraction (e.g., Blandford--Znajek) dominates
\citep{2009MNRAS.397.1153K}. Type~II collapsars thus
\textcolor{black}{provide a natural explanation for} underluminous GRBs, precursor
activity, and gradually rising prompt-emission light curves.

At this stage, the \textcolor{black}{time-dependent mass-supply} profile is adopted
\textcolor{black}{as a phenomenological description of the central-engine
evolution, without attempting a detailed classification of individual GRBs}.
A quantitative classification based on detailed light-curve morphology,
energetics, and spectral evolution will be developed in future work.

\textcolor{black}{
In Type~II collapsars, the mass supply to the central engine is dominated by the
fallback of marginally bound stellar material, which naturally exhibits a
well-defined early-time onset followed by a late-time decay that approaches a
power-law scaling $\propto t^{-5/3}$ in self-similar fallback solutions
\citep{2001ApJ...550..410M,2013ApJ...767L..36W}. To capture this behavior in a
flexible yet physically motivated manner, we adopt a smooth,
fallback-motivated \emph{time-dependent mass-supply} profile to parameterize the
engine feeding history in Type~II systems.
}

\textcolor{black}{
Specifically, we employ the following functional form for the mass-supply rate:
}

\begin{equation}\label{eq:fb}
\dot{M}(t) = \dot{M}_p
\left[
\frac{1}{2}
\left( \frac{t - t_0}{t_p - t_0} \right)^{-s/2}
+
\frac{1}{2}
\left( \frac{t - t_0}{t_p - t_0} \right)^{5s/3}
\right]^{-1/s},
\end{equation}

\textcolor{black}{
where $\dot{M}_p$ is the peak mass-supply rate, $t_0$ marks the onset of engine
activity, $t_p$ denotes the time of peak mass delivery, and $s$ controls the
smoothness of the transition between the rising (onset) and decaying phases.
This prescription is intended to represent a \emph{generic, smoothly evolving
mass-supply history} with a well-defined onset, peak, and decay, rather than a
unique or universal fallback solution.
}

\textcolor{black}{
Classical fallback provides one physically motivated realization of such a
mass-supply history, particularly relevant for delayed-collapse (Type~II)
collapsars, but the AMIS framework itself does not require fallback to dominate
the mass budget in all realizations. Instead, the adopted form serves as a
convenient and physically grounded parameterization of time-modulated engine
feeding. Such configurations may be especially relevant for low-luminosity GRBs,
including GRB~980425, where mildly relativistic ejecta and internal-shock
dissipation have been invoked to explain the prompt emission within a collapsar
context \citep[e.g.,][]{2007A&A...465....1D}.
}

\textcolor{black}{
For Type~I collapsars, the accretion is driven by prompt infall through a
hyperaccreting disk capable of powering a relativistic jet
\citep{1999ApJ...524..262M,2001ApJ...557..949N}.
 The detailed temporal evolution of the accretion rate in such
prompt-collapse scenarios remains uncertain and depends sensitively on the
progenitor structure and angular-momentum profile. It is nevertheless expected 
to be more strongly peaked and shorter-lived than the
extended fallback accretion characteristic of Type~II collapsars, with a rapid
decline of the \emph{engine power} once the dense stellar core has been exhausted
\citep{2001ApJ...550..410M,2025ApJ...992L...3G}, rather than a
$t^{-5/3}$ fallback-dominated phase \citep{1989ApJ...346..847C}.
}

\textcolor{black}{
In this work, we nevertheless employ the same functional form
(Equation~\ref{eq:fb}) for Type~I collapsars as a \emph{parametric modeling
convenience}. This choice allows us to explore, in a unified manner, how a
time-modulated engine power imprints global trends on the prompt emission,
while explicitly recognizing that the late-time $t^{-5/3}$ behavior is not
physically motivated for prompt-collapse events. Indeed, a more rapidly declining engine power is more effective to producing
the sharply decaying ``exponential'' tail commonly observed in classical FRED
light curves \citep{1996ApJ...459..393N,2005ApJ...627..324N}.
In practice, for Type~I collapsars the model parameters can be chosen
to encode a more intense and short-lived accretion episode than
in Type~II systems. Future multidimensional simulations of collapsar accretion and jet launching
will be required to replace this prescription with a more
physically realistic description of $\dot{M}(t)$ in prompt-collapse systems. 
}

The instantaneous jet (engine) power is parameterized as
\begin{equation}
\dot{E}_{\rm jet}(t) = \eta\, \dot{M}(t) c^2,
\end{equation}
where \(\eta\), \textcolor{black}{which varies} in the range
\(\sim 10^{-3} - 0.1\), accounts for the efficiency of converting rest-mass
energy into jet power, encompassing neutrino, MHD, or spin-extraction processes
\citep{1999ApJ...518..356P,2009MNRAS.397.1153K}.
\textcolor{black}{For the parameter ranges
primarily explored in this work, the implied peak accretion rates,
$\dot{M}_p \sim 10^{-2}$--$10^{-1}\,M_\odot\,\mathrm{s^{-1}}$, are consistent with
hyperaccreting inner flows commonly inferred for Type~I collapsar GRB engines,
depending on the disk structure and accretion geometry. Lower peak accretion
rates, $\dot{M}_p \sim 10^{-3}$--$10^{-2}\,M_\odot\,\mathrm{s^{-1}}$, may also be
realized in Type~II collapsar scenarios, particularly when combined with higher
radiative or internal-shock efficiencies. 
}

\textcolor{black}{Accordingly, the time-dependent mass supply to the central engine is assumed,
when accretion conditions permit, to power a relativistic outflow whose injected
energy is partially converted into non-thermal radiation via internal
dissipation processes such as shocks or magnetic reconnection.} The observable prompt gamma-ray luminosity is then parameterized as
\begin{equation}
L_\gamma(t) = \epsilon_{\rm rad}\, \dot{E}_{\rm jet}(t)
= \epsilon_{\rm rad}\, \eta\, \dot{M}(t)\, c^2 ,
\end{equation}
where $\epsilon_{\rm rad}$ denotes the radiative efficiency, i.e.\ the fraction
of jet power converted into gamma-ray emission
\citep[e.g.,][]{Meszaros2006,2015PhR...561....1K}.

For typical GRB parameters, a substantial fraction of this prompt emission
\textcolor{black}{is expected to} arise from synchrotron radiation of relativistic
electrons accelerated within internal dissipation regions of the jet. The
characteristic photon energy can be estimated as
\begin{equation}
E_{\rm syn} \propto
\Gamma~B'~\gamma_e^2,
\end{equation}
where \(\Gamma\) is the bulk Lorentz factor of the jet, \(B'\) is the comoving
magnetic field, and \(\gamma_e\) is the Lorentz factor of the radiating
electrons. For typical values (\(\Gamma \gtrsim 300\),
\(B' \sim 10^4--10^5~{\rm G}\), \(\gamma_e \sim 10^2--10^4\)), synchrotron
emission \textcolor{black}{can naturally produce} photons in the
100--1000~keV range, consistent with observed prompt GRB spectra
\citep[e.g.,][]{2015PhR...561....1K, Meszaros2006}.

\textcolor{black}{
Throughout this work, the term ``\textcolor{black}{Accretion}-Modulated'' is used in a
phenomenological sense to denote a time-dependent mass-supply history regulating
the engine power. Classical fallback provides one physically motivated realization
of such modulation, but the \textcolor{black}{A}MIS framework does not require fallback to
dominate the accretion flow.
}

\textcolor{black}{\section{Accretion-Modulated Internal Shocks (AMIS)}\label{sec:FMIS}}
\subsection{Kobayashi-Piran-Sari (KPS) Collision Dynamics for Two/Multiple Shells}

Internal shocks (ISs) were first modeled as collisions of discrete shells \citep{1997ApJ...490...92K}. 
In its multi–shell scenario, the succession of shells emanate from the inner engine. Therefore, the long-term light curve reflects the central-engine activity, while local fluctuations arise from stochastic variations in shell masses and Lorentz factors. If the engine drives a secular trend in shell ejection—such as one set by a time-dependent fallback accretion rate—the same trend is expected to appear in the IS light curve because observed pulse times map approximately one-to-one onto 
engine ejection times. 

A nonuniform relativistic wind may therefore be represented as a sequence of shells, 
with two–shell collisions as the elementary dissipation process. 
A faster shell ($r$) catches a slower one ($s$), merging into a new shell ($m$); 
energy–momentum conservation gives the Lorentz factor of the merged shell \citep{1997ApJ...490...92K}. 
Superposed pulses from successive collisions reproduce the observed temporal variability. For two colliding cold shells,
\begin{eqnarray}
\label{gammam}
\gamma_m &=& 
  \left[\frac{m_r\gamma_r+m_s\gamma_s}{m_r/\gamma_r+m_s/\gamma_s}\right]^{1/2},\\[4pt]
E_{\rm int} &=& 
  m_r c^2(\gamma_r-\gamma_m) + m_s c^2(\gamma_s-\gamma_m),
  \label{Eint}
\end{eqnarray}
where $\gamma_m$ is the post–collision bulk Lorentz factor and $E_{\rm int}$ is the internal (dissipated) energy. 
The corresponding radiative efficiency is 

\begin{equation}
    \label{efficiency}
    \epsilon = 1 - (m_r+m_s)\gamma_m/(m_r\gamma_r+m_s\gamma_s). 
\end{equation}

The Lorentz factors of the forward and reverse shocks are given by \citep{1997ApJ...490...92K}:
\begin{eqnarray}
\gamma_{\rm fs} \simeq \gamma_m \sqrt{\frac{1 + 2\gamma_m/\gamma_s}{2 + \gamma_m/\gamma_s}},\label{fs} \,
\gamma_{\rm rs} \simeq \gamma_m \sqrt{\frac{1 + 2\gamma_m/\gamma_r}{2 + \gamma_m/\gamma_r}},
\label{rs}
\end{eqnarray}
with corresponding velocities $\beta_{\rm fs} = \sqrt{1 - \gamma_{\rm fs}^{-2}}$ and $\beta_{\rm rs} = \sqrt{1 - \gamma_{\rm rs}^{-2}}$.

The merged shell width is calculated as \citep{1997ApJ...490...92K}:
\begin{equation}\label{lm}
l_m = l_s \frac{\beta_{\rm fs} - \beta_m}{\beta_{\rm fs} - \beta_s} + l_r \frac{\beta_m - \beta_{\rm rs}}{\beta_r - \beta_{\rm rs}},
\end{equation}
where $l_s$ and $l_r$ are the widths of the slow and rapid shells, and $\beta_m$, $\beta_s$, $\beta_r$ are the velocities of the merged, slow, and rapid shells respectively.

The reverse–shock crossing (emission) time and the curvature (angular) timescale are estimate as
\begin{eqnarray}
\delta t_e = \frac{l_r}{c(\beta_r-\beta_{\rm rs})},~\,
\tau_{\rm ang} = \frac{R}{2\gamma_m^2 c},
\end{eqnarray}
where $l_r$ is the shell width in the lab frame, $\beta_{\rm rs}$ is the velocity of the reverse shock, 
and $R$ is the collision radius. 

The analytic pulse shape proposed by \citet{1997ApJ...490...92K} is adopted as

\begin{equation}\label{kps}
L_{\rm KPS}(t)=h 
\begin{cases}
1-\dfrac{1}{(1+t/\tau_{\rm ang})^2}, & 0<t<\delta t_e/(2\gamma_m^2), \\[6pt]\\
\dfrac{1}{(1+t/\tau_{\rm ang}-\Delta)^2}-\\\dfrac{1}{(1+t/\tau_{\rm ang})^2}, & t>\delta t_e/(2\gamma_m^2),
\end{cases}
\end{equation}
where $\Delta\equiv c\delta t_e/R$ and $h=(2\gamma_m^2/\delta t_e)E_{\rm int}$ ensures $\int L_{\rm KPS}(t)\,dt=E_{\rm int}$. 

\begin{figure}
    \centering
    \includegraphics[width=0.48\textwidth]{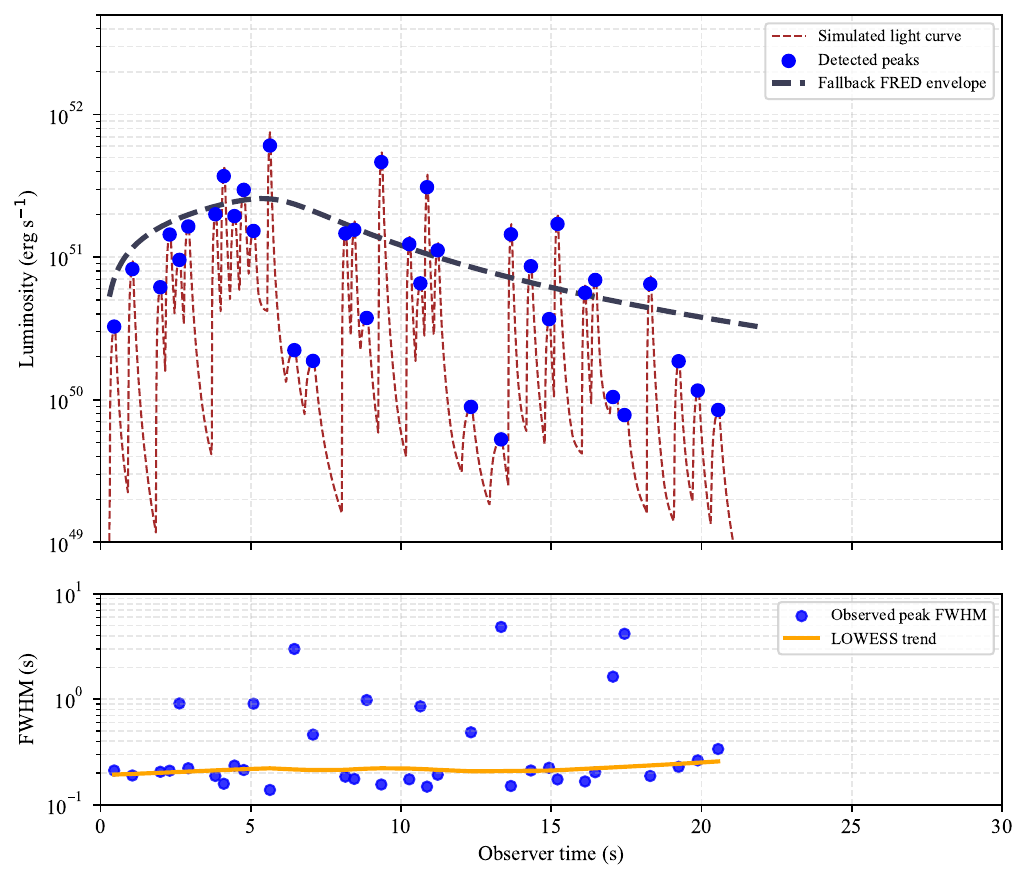} 
 \caption{Simulated GRB prompt-emission light curves from an
\textcolor{black}{Accretion}-Modulated internal-shock (\textcolor{black}{A}MIS) model.
\textbf{Mass-driven scenario.} The fast-rise, slow-decay shape of the
envelope is imposed by \textcolor{black}{a time-dependent, accretion-regulated
mass-supply history of the central engine}, associated with stellar collapse
\textcolor{black}{and accretion-driven mass delivery}, which modulates the ejected
shell masses ($m_0 = 3\times10^{28}$ g, $t_0=0.1$ s, $t_p=6$ s, $s=5$, and lab-frame
widths $l_i \in [10^{9},2\times10^{9}]\,\mathrm{cm}$) at roughly constant ejection
intervals. Stochastic variations in shell Lorentz factors
($\Gamma_i \in [400,800]$) generate internal shocks, producing the fine-scale,
multi-peaked structure typical of GRB light curves. The simulation yields a
total dissipated energy $E_{\rm diss} \simeq 1.036\times10^{52}$ erg, a total
initial kinetic energy $E_{\rm kin} \simeq 5.542\times10^{53}$ erg, and a
time-averaged radiative efficiency $\epsilon \simeq 0.019$.
Pulse FWHMs are measured from the observed peaks of the smoothed light curve,
ensuring that closely spaced or merged collisions are treated as single pulses.
The resulting FWHMs remain approximately constant, reflecting the nearly uniform
hydrodynamic and angular timescales associated with fixed ejection intervals. A
locally weighted smoothing (LOWESS) of the individual pulse widths highlights
only minor fluctuations around this overall flat trend.
}
\label{fig:FMIS_lightcurves}
\end{figure}

\begin{figure}
    \centering
    \includegraphics[width=0.48\textwidth]{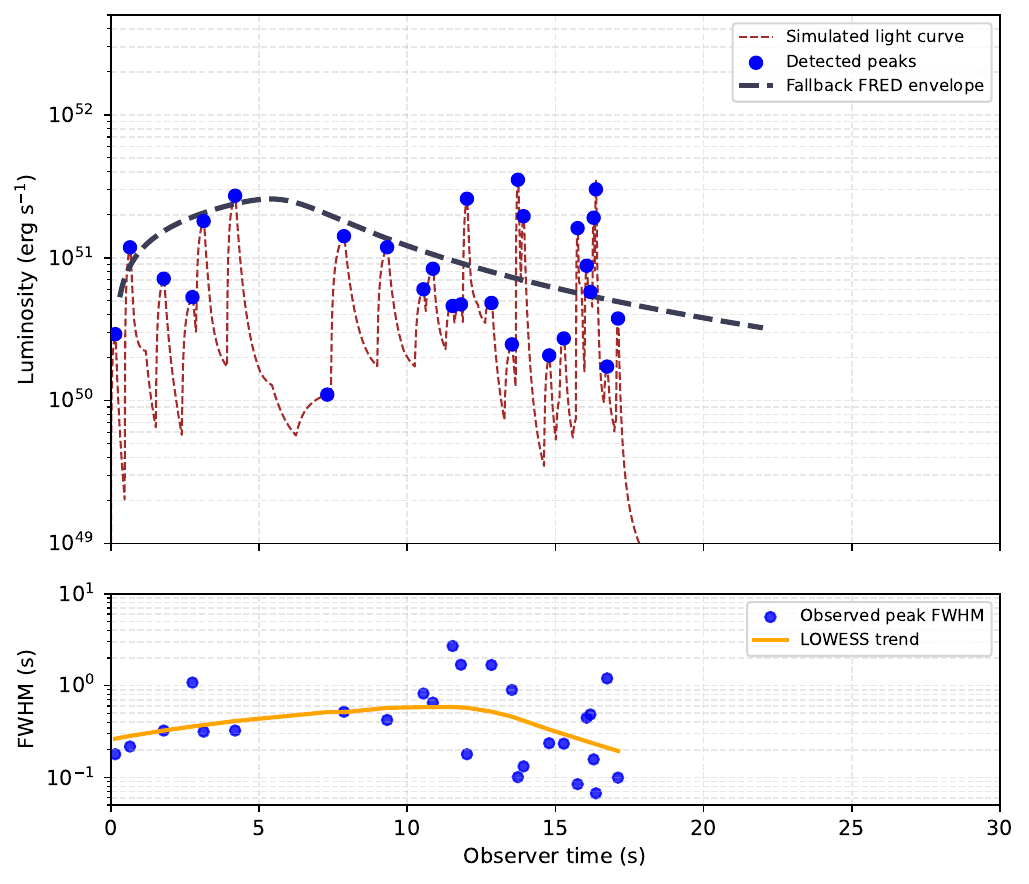}
    \caption{Simulated GRB prompt-emission light curves from an
\textcolor{black}{Accretion}-Modulated internal-shock (\textcolor{black}{A}MIS) model.
\textbf{Rate-driven scenario.} Shell ejection intervals follow the
\textcolor{black}{time-dependent, accretion-regulated mass-supply history},
while the shell masses remain nearly constant
($m_i \sim 3\times10^{28}$ g). Random fluctuations in the Lorentz factors cause
shells to collide at irregular intervals, generating internal shocks that
imprint variability on the light curve. The simulation yields a total
dissipated energy $E_{\rm diss} \simeq 1.526\times10^{52}$ erg, a total kinetic
energy $E_{\rm kin} \simeq 1.062\times10^{54}$ erg, and a time-averaged
radiative efficiency $\epsilon \simeq 0.014$.
Pulse FWHMs evolve with the \textcolor{black}{mass-supply rate}: pulses are
narrower and brighter during phases of lower mass supply and broader and
dimmer during phases of higher mass supply. The trend is derived using locally
weighted smoothing over individual pulse widths to highlight the overall
temporal evolution.
}

    \label{fig:FMIS_lightcurves1}
\end{figure}

\subsection{Multiple-Shell Collisions}

In this scenario, we assume that the central engine releases a sequence of shells in a single, short-lived activity envelope episode without long quiescent intervals. At the lab-frame time $\tilde t = 0$, when the last shell is emitted, the inner edges of the shells are located at
\begin{equation}
R_i(0) = -\,\tilde t_i \beta_i c, 
\qquad \beta_i = \sqrt{1-\gamma_i^{-2}},
\end{equation}
where $\tilde t_i$ is the shifted ejection time such that $\tilde t_N = 0$ for the last shell. 
The separation between shells $i$ and $i+1$ is
\begin{equation}
L_i = R_i - R_{i+1} - l_{i+1}.
\end{equation}
Following \citet{1997ApJ...490...92K}, the collision time between two adjacent shells 
(inner faster, outer slower) is
\begin{equation}
\delta t_{i,i+1} = \frac{L_i}{c(\beta_{i+1} - \beta_{i})},
\end{equation}
which is positive only when $\beta_{i+1} > \beta_{i}$. 

For $N>2$ shells, collisions are handled recursively. At each step, the minimum $\delta t_{i,i+1}$ is selected, the system time is advanced by that amount, 
and the colliding pair merges using the conservation relations. The complete merging procedure is: 1) Calculate merged Lorentz factor $\gamma_m$ using equation~\ref{gammam}; 2) Calculate internal energy $E_{\rm int}$ using equation~\ref{Eint}; 3) Calculate reverse and forward shock Lorentz factors $\gamma_{\rm fs}$, $\gamma_{\rm rs}$ using equations~\ref{rs}; 4) Calculate merged shell width $l_m$ using equation~\ref{lm}; 5) Update shell properties: $\gamma_m$, $m_m = m_r + m_s$, and $l_m$. 

The observed arrival time of the corresponding pulse is
\begin{equation}
t_{{\rm obs}} = (1+z)\left[t_{\rm col} - \frac{R_c(t_{\rm col})}{c}\right],
\end{equation}
where $R_c(t_{\rm col})$ is the collision radius, $z$ is the cosmological redshift of the source, and the earliest $t_{{\rm obs}}$ is set to zero.

The approximate one-to-one mapping between engine ejection times and observed pulse times 
can be verified by equating shell trajectories. 
Consider two shells ejected by the central engine: a slower shell ($s$) at $t = 0$ with velocity $\beta_s c$, 
and a faster shell ($r$) at $t = \Delta t_{\rm ej}$ with velocity $\beta_r c > \beta_s c$. 
Their positions in the lab frame are
\begin{equation}
R_s(t) = \beta_s c\, t, \qquad
R_r(t) = \beta_r c\, (t - \Delta t_{\rm ej}), \quad t > \Delta t_{\rm ej}.
\end{equation}

The collision occurs when the shells meet, $R_s(t_{\rm col}) = R_r(t_{\rm col})$, 
which gives the lab-frame collision time
\begin{equation}
t_{\rm col} = \frac{\beta_r}{\beta_r - \beta_s}\,\Delta t_{\rm ej}.
\end{equation}

The collision radius is then
\begin{equation}
R_c = R_s(t_{\rm col}) = \frac{\beta_r \beta_s}{\beta_r - \beta_s} c\, \Delta t_{\rm ej}.
\end{equation}

Accounting for the light-travel delay to the observer, the arrival time of photons is
\begin{equation}
t_{\rm obs} = t_{\rm col} - \frac{R_c}{c} = \frac{\beta_r (1 - \beta_s)}{\beta_r - \beta_s}\, \Delta t_{\rm ej}.
\end{equation}

For ultra-relativistic shells ($\beta_i \simeq 1 - 1/(2\gamma_i^2)$), this reduces to
\begin{equation}
t_{\rm obs} \simeq \frac{\gamma_r^2}{\gamma_r^2 - \gamma_s^2}\, \Delta t_{\rm ej}.
\end{equation}

Hence, for $\gamma_r \gg \gamma_s$, we have $t_{\rm obs} \approx \Delta t_{\rm ej}$,  
demonstrating that the observed pulse times approximately trace the engine ejection times. 
This justifies the one-to-one mapping between central-engine activity and observed internal-shock pulses.

\textcolor{black}{\section{Mass-Supply--Modulated Temporal Trend}\label{sec:FMIS1}}

To link the KPS internal-shock scenario with a time-dependent mass supply to the
central engine, \textcolor{black}{as an illustrative realization}, we adopt a
fallback-motivated mass-delivery history. To account for both the initial
\textcolor{black}{onset of mass delivery} and the post-peak decay, we consider a
function similar to Equation~\ref{eq:fb}, which captures the full temporal
evolution of the mass-supply history:
\begin{equation}
\dot{M}_{\textcolor{black}{\rm ms}}(t) \propto 
\begin{cases}
t^{\beta_{\rm onset}}, & t < t_{\rm peak},\\
t^{-5/3}, & t \ge t_{\rm peak},
\end{cases}
\end{equation}
with $\beta_{\rm onset} \simeq 0$--0.5 \citep{2001ApJ...550..410M}. \textcolor{black}{Here $\dot{M}_{\rm ms}(t)$ denotes the time-dependent \emph{mass-supply rate to the engine/disk}.}

\textcolor{black}{The time-dependent mass-supply rate to the central engine}
modulates the properties of the emitted shells, which can be implemented through
a time-dependent normalization factor
($h=(2\gamma_m^2/\delta t_e)E_{\rm int}$) of the KPS luminosity in
Equation~(\ref{kps}):
\begin{equation}
h_j(t_{\rm obs}) = A\,\left(\frac{t_{\rm obs}}{t_{\rm ref}}\right)^{\beta_{\rm eff}(t_{\rm obs})}
\;\propto\; \dot{M}_{\textcolor{black}{\rm ms}}(t),
\end{equation}
where
\begin{equation}
\beta_{\rm eff}(t_{\rm obs}) =
\begin{cases}
\beta_{\rm onset}, & t_{\rm obs} < t_{\rm peak},\\
-5/3, & t_{\rm obs} \ge t_{\rm peak},
\end{cases}
\end{equation}
$t_{\rm ref} = 1\,\mathrm{s}$, and $A$ is chosen to conserve energy.
\textcolor{black}{The proportionality reflects the assumption that, to leading
order, the dissipated energy per collision traces the engine mass-supply history,
with the mapping between engine time and observer time treated approximately.}

For the envelope in this \textcolor{black}{A}MIS framework, we specifically refer to the broad,
smooth luminosity evolution of GRBs whose prompt emission profiles exhibit a
single fast-rise–exponential-decay (FRED) segment. These events provide a clean
testbed for isolating \textcolor{black}{mass-supply–modulated} physics. In
multi-pulse events, each sub-envelope may correspond to a distinct
\textcolor{black}{episode of enhanced mass delivery or modulation of the engine
feeding history}, giving rise to compound FRED-like structures similar to those
observed in several BATSE and GBM bursts \citep{2014ApJ...783...88H}. Since this
work represents the first introduction of the \textcolor{black}{A}MIS framework, we focus on the
simplest single-envelope configurations; more complex cases involving multiple
\textcolor{black}{mass-supply modulation episodes} will be investigated in
forthcoming studies.

In fact, single-pulse, FRED-like GRBs constitute a significant subset of the
long-GRB population. Early BATSE analyses already highlighted the prevalence of
such smooth FRED profiles among bright bursts
\citep{1996ApJ...459..393N}, and a more recent systematic study of 2710 Fermi GRBs
based on \textcolor{black}{objective automated criteria} found that almost half of
the long bursts can be classified as \textcolor{black}{predominantly} single-peaked
\citep{Hintze2022}.

It is worth noting that in many \textcolor{black}{Type II collapsar} scenarios the late-time
\textcolor{black}{mass-supply history to the central engine} can decay as a power
law, typically with index $\sim -5/3$ after its peak time, as expected for the
self-similar fallback of marginally bound ejecta in physically motivated cases.
Moreover, as shown by \citet{2000ApJ...529L..13R,2002ApJ...566..210R}, in a
substantial fraction of GRBs the late-time portion of the photon flux curves is
better described by a power-law decay, $N(t) \propto t^{-n}$, with $n$ between
about 1 and 2. This behavior arises when both the hardness--intensity correlation
(HIC) and the hardness--fluence correlation (HFC) hold, producing a natural
transition from exponential-like early decay to power-law tails in the photon
flux. Within the \textcolor{black}{A}MIS framework, this suggests that even if individual
pulses begin with an approximately exponential decay, the cumulative emission
from successive internal shocks, driven by an engine whose injection is
\textcolor{black}{regulated by a declining mass-supply history}, can imprint a
late-time power-law decay in $N(t)$, and consequently in the \textcolor{black}{observed} flux,
consistent with the empirical findings of
\citet{2000ApJ...529L..13R,2002ApJ...566..210R,2012ApJ...752..132P}.

\subsection{Mass-Driven vs. Rate-Driven Scenarios}

We consider two distinct ways in which a \textcolor{black}{time-dependent
mass-supply history to the central engine} can modulate the prompt emission of
gamma-ray bursts. These two channels are representatives to capture the whole picture. In a physical setting, complex and complicated  \textcolor{black}{forms of mass-supply modulation} are expected.

In the \textit{mass-driven scenario}, shells are emitted at roughly constant
intervals, $\Delta t_{\rm ej} \simeq \Delta t_0$, while the mass of each
shell follows the \textcolor{black}{time-dependent mass-supply history}:
\begin{equation}
m_i \simeq \textcolor{black}{\dot{M}_{\rm ms}}(t_i)\,\Delta t_0
\propto t_i^{\beta_{\rm eff}(t_i)}.
\end{equation}
Consequently, before the peak of the \textcolor{black}{mass-supply rate}, pulse
amplitudes grow with time ($\beta_{\rm eff}>0$), while after the peak they decay
with $\beta_{\rm eff} \simeq -5/3$. With stochastic Lorentz-factor fluctuations
among the ejected shells, the dissipated energy approximately tracks the
underlying shell-mass trend ($E_{\rm int} \propto m_i$), while the hydrodynamic
and angular timescales that determine the pulse FWHM follow only the mild fluctuations of the Lorentz factors and
therefore maintain an overall near-constant trend across collisions.
This leads to mass-driven pulses that exhibit  energy
growth during the rising phase and decay after the peak \textcolor{black}{of the mass-supply
history}, while their FWHM remain approximately constant. Such behavior is in
line with observational studies indicating that pulse widths in long GRBs show
little or no systematic temporal evolution
\citep{2000ApJ...539..712R}.

Our numerical simulation of $N_{\rm shells}=70$ shells with initial masses
modulated by a \textcolor{black}{smooth FRED-like mass-supply envelope}
($m_0 = 3\times10^{28}$ g, $t_0=0.1$ s, $t_p=6$ s, $s=5$, and lab-frame widths
$l_i \in [10^{9},2\times10^{9}]\,\mathrm{cm}$) and stochastic Lorentz factors in
the range $\Gamma_i \in [400,800]$ produces a \textcolor{black}{prompt-emission} light curve
consistent with \textcolor{black}{A}MIS expectations; see Figure~\ref{fig:FMIS_lightcurves}.
The simulation yields a total dissipated energy
$E_{\rm diss} \simeq 1.036\times10^{52}$ erg and a total initial kinetic energy
$E_{\rm kin} \simeq 5.542\times10^{53}$ erg, corresponding to a time-averaged
radiative efficiency $\epsilon \simeq 0.019$.

\textcolor{black}{
In our simulations the shells are not sorted by Lorentz factor, and no
\emph{a~priori} ordering is imposed. Each shell is assigned an ejection time
according to the prescribed engine sequence, and its Lorentz factor is drawn
independently from a uniform (linear) distribution within the stated range.
Internal shocks then arise self-consistently when a faster shell emitted at a
later time catches up with a slower shell emitted earlier. Different random
realizations change the detailed pulse morphology (e.g., exact peak times and
amplitudes), as expected in stochastic internal-shock models, but do not alter
the qualitative \textcolor{black}{A}MIS trends discussed here.
}

\textcolor{black}{
The above values are physically consistent with a hyperaccreting collapsar
engine, representative of a Type~I realization, in which the time-dependent mass
supply to the central engine is regulated primarily by collapse-driven mass
delivery, with possible contributions from delayed feeding. Assuming a typical
accretion-to-outflow efficiency $\eta_{\rm acc} \sim 0.1$, the implied mass
supplied to the engine is $M_{\rm sup} \sim E_{\rm kin}/(\eta_{\rm acc}
c^2) \sim 3\,M_\odot$, corresponding to an effective accretion rate of
$\dot M \sim 0.15\,M_\odot\,\mathrm{s^{-1}}$ over the first $\sim 20$~s. These
values lie well within the hyperaccretion regime expected during the prompt
phase of Type~I collapsars.
}

\textcolor{black}{
Furthermore, for mildly relativistic outflows relevant to low-luminosity GRBs,
shell Lorentz factors in the range $\Gamma\sim 10$--$50$ can still produce
internal-shock dissipation efficiencies of order $\epsilon_{\rm IS}\sim 0.1$
(10\%) when the shell-to-shell Lorentz-factor contrasts are moderate \citep{1997ApJ...490...92K, 2007A&A...465....1D}.  With an
engine (accretion-to-jet) conversion efficiency $\eta_{\rm acc}\sim 0.1$, this
level of dissipation implies that the required mass-supply rate can be as low as
$\dot{M}\sim 2\times 10^{-3}\,M_\odot\,\mathrm{s^{-1}}$ to account for the prompt
energetics, making the scenario naturally compatible with Type~II collapsars
sustained by longer-lasting, \emph{fallback-motivated mass-supply histories} at
relatively modest instantaneous rates.
}

Moreover, our simulation shows that the number of internal shock collisions
exceeds the number of \textit{observed} pulses. In practice, several closely
spaced collisions may merge into a single pulse in the light curve, and the
FWHM of individual pulses cannot be inferred directly from the collision list.
Therefore, to measure pulse widths from the observed light curve itself, we
first generated a smoothed version of the light curve using a Gaussian kernel
($\sigma=2$ bins), which suppresses the high-frequency variability associated
with individual collisions while preserving the broader pulse morphology.
Local maxima in the smoothed light curve were then identified, and for each peak
the left and right half-maximum crossing times were located by stepping outward
from the peak and linearly interpolating between the grid points.

Furthermore, to observe the overall evolution of FWHMs with time, we applied a
locally weighted regression (LOWESS; \citealt{Cleveland1979}), which provides a
non-parametric estimate of the FWHM--$t_{\rm pk}$ relation. As shown in
Figure~\ref{fig:FMIS_lightcurves}, the trend clearly shows mild fluctuations
around a nearly constant value, consistent with the expectation in the
\textcolor{black}{mass-driven AMIS scenario}. To assess the robustness of the
result, further collisions with smaller peaks were added, and the FWHM trend
remained the same.

\subsection{Rate-Driven Scenario}

In the rate-driven limit, the \textcolor{black}{time-dependent mass-supply history
to the central engine} regulates the shell \emph{ejection interval} \textcolor{black}{within the
AMIS framework}:
\begin{equation}
\Delta t_i \propto \textcolor{black}{\frac{1}{\dot{M}_{\rm ms}(t_i)}},
\qquad
L_i = c\,\Delta t_i ,
\end{equation}
while the shell masses remain nearly constant ($m_i\simeq m_0$). In the KPS model
of internal shocks \citep{1997ApJ...490...92K}, the separation of the shell $L_i$
and the width of the shell $\ell_i$ are independent engine parameters. $L_i$
represents the waiting time between consecutive ejections, and $\ell_i$ is set
by the duration of the engine's \textit{ON} phase.

Therefore, unlike the original KPS treatment, we adopt a relation in which the
duration of engine activity during each ejection episode is tied to the same
\textcolor{black}{mass-supply--controlled} timescale \textcolor{black}{within the A}MIS framework,
\begin{equation}
\ell_i = c\,\delta t_{\rm ON}(t_i),
\qquad
\delta t_{\rm ON}(t_i) \propto \Delta t_i ,
\end{equation}
so that periods of high \textcolor{black}{mass-supply rate} produce long
engine-activity episodes and therefore thick shells, whereas a declining
\textcolor{black}{mass-supply history} leads to progressively thinner shells.

With fixed shell masses, the dissipated energy in each collision is
approximately constant, $E_{\rm int}\propto m_0$, with random variations tied to
the Lorentz-factor contrasts. Therefore, the resulting pulse properties are
determined by the radial shock-crossing time, which scales with the shell
thickness. When $\delta t_e \gg \tau_{\rm ang}$, the peak luminosity follows the
KPS scaling,
\begin{equation}
L_{\rm pk}(t_i)
\;\propto\;
\frac{E_{\rm int}}{\delta t_e(t_i)}
\;\propto\;
\frac{1}{\ell_i(t_i)} ,
\end{equation}
implying that thick shells produced during phases of
\textcolor{black}{enhanced mass-supply rate} yield broad, low-luminosity pulses,
while the thinner shells ejected as the
\textcolor{black}{mass-supply history declines} produce narrower and brighter
pulses. When $\delta t_e \ll \tau_{\rm ang}$, the angular timescale dominates and
both the pulse width and $L_{\rm pk}$ saturate.

Our simulation of 70 shells with $m_i\simeq 3\times10^{28}$\,g and
$\Gamma_i\in[400,800]$ yields a dissipated energy
$E_{\rm diss}\simeq 1.53\times10^{52}$\,erg, out of an initial kinetic energy
$E_{\rm kin}\simeq 1.06\times10^{54}$\,erg, corresponding to a time-averaged
radiative efficiency $\epsilon\simeq0.014$. The inverse relation between pulse
width and peak luminosity emerges naturally, producing the width--luminosity
evolution visible in Fig.~\ref{fig:FMIS_lightcurves1}. For a \textcolor{black}{narrower}
Lorentz-factor distribution ($\Gamma\sim10$--$50$),  the effective
\textcolor{black}{mass-supply requirement reduces} toward values
\textcolor{black}{compatible with longer-lasting, fallback-motivated feeding
histories}, as commonly discussed for Type~II collapsar realizations.

To summarize, in this strict rate-driven formulation, the
\textcolor{black}{time-dependent mass-supply history} modulates the ejection timing
and shell thicknesses, and therefore regulates pulse widths. However, the pulse
amplitudes do not directly follow the \textcolor{black}{mass-supply} envelope since
$E_{\rm int}$ remains almost fixed. A more physical generalization, in which both
$\Delta t_i$ and the injected energy (or shell mass) scale with
\textcolor{black}{the instantaneous mass-supply rate}, would imprint the same
\textcolor{black}{mass-supply–modulated} profile on both the pulse widths and
amplitudes, leading naturally to a smooth FRED-like envelope.

\subsection{Global Envelope Characterization via Norris FRED Fitting}
\label{sec:norris_fit}

To quantify the large-scale temporal behavior of the \textcolor{black}{mass-driven,
mass-supply--modulated} \textcolor{black}{A}MIS light curve, as an indicative example, we fit the
simulated luminosity envelope using the analytic FRED form of
\citet{1996ApJ...459..393N,2005ApJ...627..324N}.

The pulse shape is \citep{2005ApJ...627..324N,2014ApJ...783...88H}
\begin{equation}
F(t) = A \exp\!\left[-\frac{\tau_1}{t - t_{\rm start}} -
\frac{t - t_{\rm start}}{\tau_2}\right],
\qquad t>t_{\rm start},
\end{equation}
where $A$ is the pulse amplitude, $t_{\rm start}$ the effective onset time, and
$\tau_1$ and $\tau_2$ the rise and decay timescales, respectively.

\begin{figure}[t]
\centering
\includegraphics[width=0.95\linewidth]{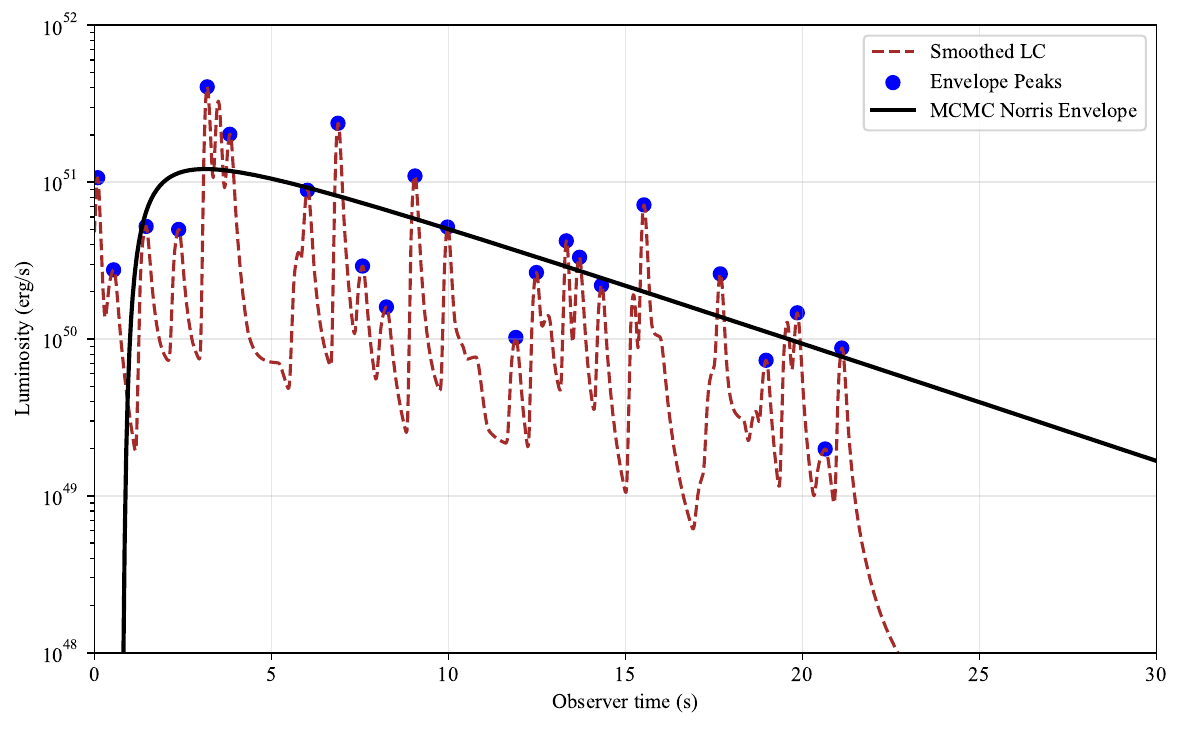}
\includegraphics[width=0.95\linewidth]{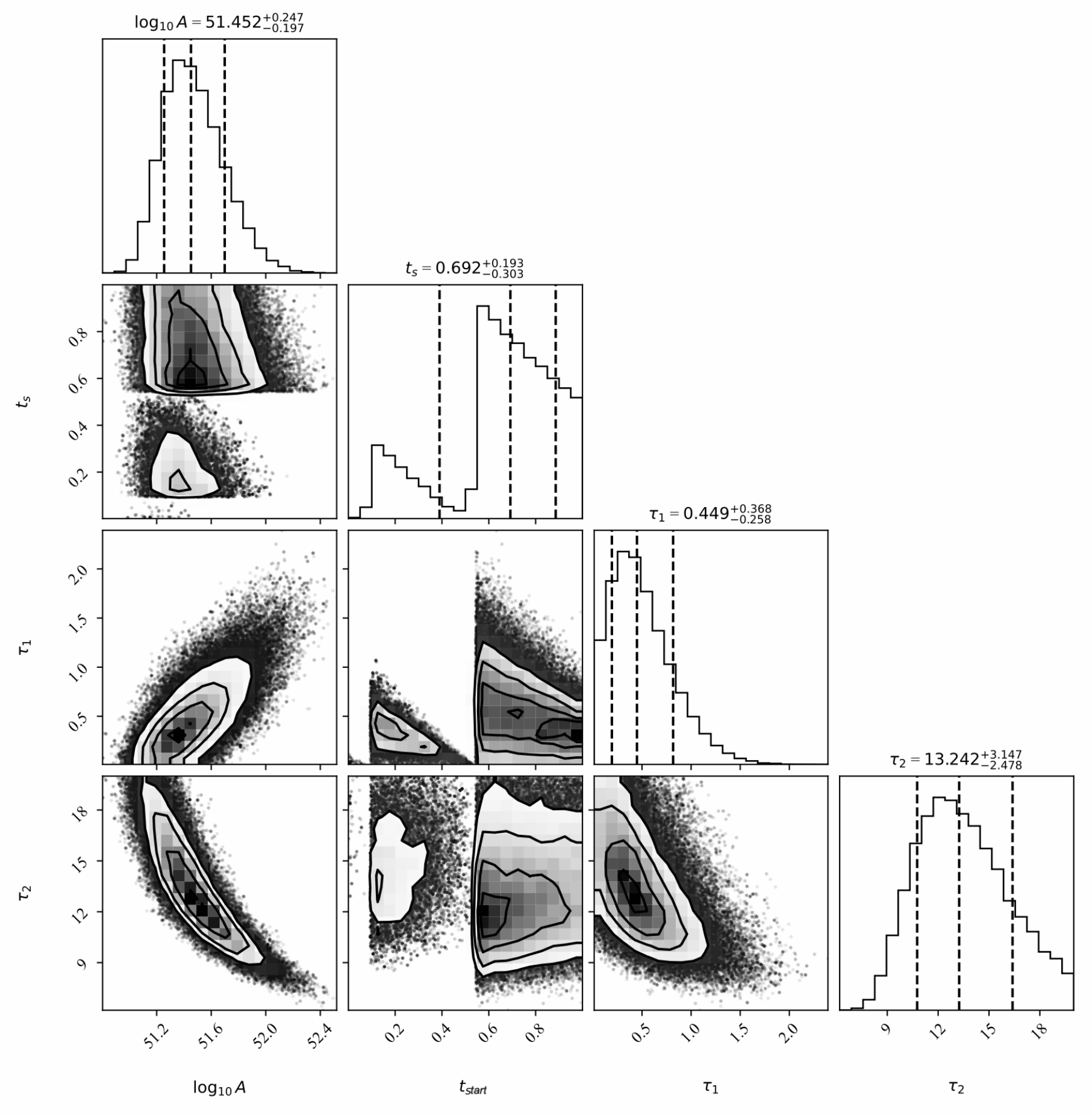}
\caption{
\citet{2005ApJ...627..324N} pulse fitting of the mass-driven \textcolor{black}{A}MIS luminosity
envelope and the corresponding MCMC parameters. The blue points represent the
most significant small pulse peaks used for the fit, the dashed curve shows the
\textcolor{black}{A}MIS simulated luminosity, and the solid black line shows the Norris FRED
profile. \textcolor{black}{The full MCMC posterior distributions are shown in the
corner plot, while the prior ranges and the posterior median values with 68\%
credible intervals are summarized in Table~\ref{tab:mcmc_priors_post}. Best-fit
parameters (defined as the posterior median values with 68\% credible
intervals) are $A = (2.83^{+2.18}_{-1.03})\times10^{51}$ erg s$^{-1}$,
$t_{\rm start} = 0.692^{+0.193}_{-0.303}$ s, $\tau_1 =
0.449^{+0.368}_{-0.258}$ s, and $\tau_2 = 13.242^{+3.147}_{-2.478}$ s. The
inferred FRED peak time, $t_{\rm pk}=t_{\rm start}+\sqrt{\tau_1\tau_2} =
3.1\pm0.9$ s, coincides with the \textcolor{black}{characteristic peak timescale
of the imposed mass-supply envelope} ($t_p=3.5$ s).}
}
\label{fig:FRED_fit_massdriven}
\end{figure}

To perform a more reliable fit, we identified the significant envelope peaks in a
Gaussian-smoothed version of the light curve and fitted only these points in
logarithmic luminosity space. Figure~\ref{fig:FRED_fit_massdriven} shows the fit
together with the simulated \textcolor{black}{A}MIS envelope, with the posterior distributions of
the MCMC fitting parameters $(\log A, t_{\rm start}, \tau_1, \tau_2)$ and their
uncertainties. The \textcolor{black}{best-fit parameters (defined as posterior
median values with 68\% credible intervals)} are:
\textcolor{black}{$A = (2.83^{+2.18}_{-1.03})\times10^{51}\ {\rm erg\,s^{-1}}$},
\textcolor{black}{$t_{\rm start} = 0.692^{+0.193}_{-0.303}~{\rm s}$},
\textcolor{black}{$\tau_1 = 0.449^{+0.368}_{-0.258}~{\rm s}$},
\textcolor{black}{and $\tau_2 = 13.242^{+3.147}_{-2.478}~{\rm s}$},
all of which fall within the observed distributions of bright long-GRB pulses
\citep[e.g.,][]{1996ApJ...459..393N,2003ApJ...596..389K,2012ApJ...744..141B,2014ApJ...783...88H}.
\textcolor{black}{We adopt uniform priors within the bounds listed in
Table~\ref{tab:mcmc_priors_post}.}

For the \citet{2005ApJ...627..324N} pulse model, the pulse-peak time is given
analytically by
$t_{\rm pk} = t_{\rm start} + \sqrt{\tau_1 \tau_2}.$
Using the fitted values, we obtain $t_{\rm pk} = 3.1\pm 0.9~{\rm s}$, in agreement
with the \textcolor{black}{characteristic timescale of the imposed mass-supply
envelope in the simulation} ($t_p = 3.5$ s).

\begin{table}[t]
\centering
\caption{\textcolor{black}{Priors and posterior constraints of the Norris FRED fit
parameters derived from the MCMC analysis shown in
Fig.~\ref{fig:FRED_fit_massdriven}. The priors correspond to the uniform bounds
adopted in the MCMC sampling. Quoted uncertainties correspond to the 68\% credible intervals.}}
\label{tab:mcmc_priors_post}
\begin{tabular}{lccc}
\hline
Parameter & Prior range & Posterior median & 68\% interval \\
\hline
$A$ ($\times10^{51}$ erg s$^{-1}$)
& $[10^{-3},\,10^{4}]$
& $2.83$
& $^{+2.18}_{-1.03}$ \\
$t_{\rm start}$ (s)
& $[0,\,5.0]$
& $0.692$
& $^{+0.193}_{-0.303}$ \\
$\tau_1$ (s)
& $[0.01,\,20]$
& $0.449$
& $^{+0.368}_{-0.258}$ \\
$\tau_2$ (s)
& $[0.1,\,20]$
& $13.242$
& $^{+3.147}_{-2.478}$ \\
\hline
\end{tabular}
\end{table}

\subsection{A Note on Mass-Supply Mapping}
\label{sec:phenom_map}

In the current \textcolor{black}{A}MIS framework, we assume a
\textit{phenomenological} direct, power-law mapping between the
\textcolor{black}{time-dependent mass-supply history to the central engine},
$\textcolor{black}{\dot{M}_{\rm ms}}(t) \propto t^{\beta}$, and the modulated shell
properties. \textcolor{black}{Classical fallback accretion provides a physically
motivated realization of such a mass-supply history, particularly relevant for
Type~II collapsars.} This simplified assumption establishes the foundations of
the framework and allows exploration of its fundamental observational
signatures.

Variations in the \textcolor{black}{mass-supply rate} may introduce correlated
changes in jet luminosity, mass-loading, and variability timescales. In fact,
GRMHD simulations indicate that jet power depends non-linearly on the accretion
rate—although the precise scaling exponent is uncertain—and is influenced by
black hole spin, disk magnetic flux, and magnetization state
\citep[e.g.,][]{2011MNRAS.418L..79T,2020MNRAS.494.3656L,2022MNRAS.510.4962G}.

Future work that incorporates time-dependent GRMHD simulations of jet launching
into the \textcolor{black}{A}MIS framework could replace the
\textcolor{black}{simple power-law mass-supply prescription} with a physically
consistent relation between accretion rate and shell properties, turning this
first-order mapping into a more concrete, predictive quantity grounded in the
engine physics.

\textcolor{black}{
We emphasize that the shell properties and pulse timescales derived within the
\textcolor{black}{A}MIS framework correspond to intrinsic, source-frame dissipation
timescales set by the engine-regulated mass-supply history. In realistic
collapsar environments, these intrinsic pulse widths and luminosity trends may
be further reshaped by jet collimation, propagation inside the stellar envelope,
and viewing-angle effects, as demonstrated by multidimensional collapsar
simulations \citep[e.g.,][]{1999ApJ...524..262M,2001ApJ...550..410M}. The
\textcolor{black}{A}MIS framework is therefore intended to describe the
engine-side boundary conditions for prompt emission, rather than to replace any
sources of variability related to geometry. In this context, low-luminosity sources such as GRB~980425, which have often been
discussed in terms of off-axis viewing, can also be naturally accommodated within
the AMIS framework as intrinsically weak explosions characterized by modest
Lorentz factors and inefficient internal dissipation, without invoking an
extreme viewing-angle geometry \citep[e.g.,][]{2007A&A...465....1D}.
}

\textcolor{black}{\section{Energy-Dependent Pulse Widths in Mass-Supply--Modulated GRBs}\label{sec:energy-width}}

In the \textcolor{black}{AMIS} picture, as mentioned previously, the prompt emission
is ultimately powered by accretion onto the central object, and the jet luminosity
is assumed to scale with the instantaneous accretion rate. Once the
\textcolor{black}{mass-supply rate} reaches its maximum at $t_p$ and enters the
decaying phase, the luminosity follows the expected power--law decline with
\textcolor{black}{an approximately constant} decay index, \textcolor{black}{which we
adopt here as an indicative value, which for a fallback-modulated
realization corresponds to} $-5/3$.

Time-resolved spectral studies show that $E_{\rm p}(t)$ evolves coherently with
the instantaneous flux, either through hard--to--soft evolution or an
intensity--tracking pattern \citep[e.g.,][]{2012ApJ...756..112L,2015A&A...583A.129Y}.
Therefore, motivated by observations, the evolution of the spectral peak energy
is expected to follow the luminosity via
\begin{equation}
E_{\rm p}(t) = E_{0}\left[\frac{L_{\rm jet}(t)}{L_{0}}\right]^{\gamma_{\rm ep}},
\end{equation}
where $E_0$ and $L_0$ denote the peak energy and luminosity at $t_0$, and
$\gamma_{\rm ep}$ reflects the underlying emission physics.

Such correlations imply that the peak energy is closely related to the radiative
process and the dissipation conditions in the emission region, potentially
through variations in magnetic field strength, particle--acceleration
efficiency, or other microphysical parameters
\citep[e.g.,][and references therein]{2024ApJ...977..155M}.

The photon flux in a desired energy band $[E_1,E_2]$ is expressed as
\begin{align}
N(t;E_1,E_2) = \frac{F_E(t)\,\varphi(E_{\rm p}(t))}{E_{\rm p}(t)\,\varphi_0}, \nonumber\\
\varphi(E_{\rm p}) = \int_{E_1/E_{\rm p}}^{E_2/E_{\rm p}} B(x)\,dx,
\end{align}
where $F_E(t)$ is the bolometric energy flux,
$\varphi_0 = \int_0^\infty x\,B(x)\,dx$ is a normalization constant, and $B(x)$
is the standard normalized Band function \citep{1993ApJ...413..281B}. If the Band
indices $\alpha$ and $\beta$ are treated as constants, then
$\varphi(E_{\rm p}(t))$ acts merely as a multiplicative factor for each band, and
all different energy bands exhibit identical temporal pulse widths.

Observations and modeling show that the Band indices are effectively
energy-dependent: the low-energy slope $\alpha$ correlates with the energy
dependence of GRB pulse widths \citep{2012ApJ...752..132P}, and analyses of
individual GBM bursts demonstrate that the fitted $\alpha$ changes when the
low-energy bandpass is varied \citep{2013A&A...550A.102T}. Simulations further
confirm that, even for a fixed intrinsic spectrum, the recovered $\alpha$
depends on the instrumental bandpass because different portions of the curved
spectrum are sampled \citep{2015MNRAS.451.1511B}.

To account for this, we adopt a simple phenomenological model in which the
low-energy index depends on photon energy,
\begin{equation}
\alpha(E) = \alpha_0 + k \log \frac{E}{E_{\rm ref}},
\end{equation}
where $\alpha_0$ is the index at a reference energy $E_{\rm ref}$, and $k$ is a
positive constant, consistent with observations. Substituting this into the
asymptotic low-energy regime ($E < E_p$), the photon flux in a given band reads
\begin{align}
N_{\rm low}(t;E) \;\propto\; L_{\rm jet}(t)\,
\left[E_{\rm p}(t)\right]^{-(2+\alpha(E))} \nonumber\\
\;\propto\;
\left(\frac{t}{t_0}\right)^{-\frac{5}{3}\left[1 -
\gamma_{\rm ep}\,(2+\alpha(E))\right]}.
\end{align}

Because $\alpha(E)$ increases at lower energies, the effective decay index is
smaller for low-energy bands, producing broader pulses, whereas higher-energy
bands with smaller $\alpha(E)$ decay faster and result in narrower pulses. A
similar argument applies to the high-energy tail through $\beta(E)$.

This energy-dependent behavior can reproduce the observed scaling of pulse width
with energy, $w(E)\propto E^{-0.4}$ \citep[e.g.,][]{1996ApJ...459..393N,2015A&A...583A.129Y}.
The coupling parameter $\gamma_{\rm ep}$ controls how the evolution of the peak
energy modulates the decay, but reproducing the observed scaling requires
adjusting the phenomenological parameters $\alpha_0$, $k$, and
$\gamma_{\rm ep}$. Future iterations of the \textcolor{black}{A}MIS model should therefore
explicitly include energy- and possibly time-dependence of the Band indices to
self-consistently model the temporal evolution and pulse-width scaling of GRB
emission.

\section{Discussions and Conclusions}\label{sec:conclusions}

This work presents a framework in which the time dependence of the
\textcolor{black}{engine mass-supply history, arising from collapse-driven mass
delivery and delayed feeding of bound material,} shapes the prompt emission of
long GRBs. In this picture, the broad temporal envelope of the burst reflects
the smooth rise and decay of the \textcolor{black}{effective mass-supply rate to
the central engine}, while the rapid variability is produced by stochastic
fluctuations in the Lorentz factors and the masses or ejection intervals of
individual shells that collide through internal shocks. Within the standard
collapsar scenario, Type~I collapsars, with rapid black-hole formation and high
early accretion rates, and Type~II collapsars—with longer-lasting, fallback-like
feeding histories, can both produce such behavior. The \textcolor{black}{A}MIS calculations shown
here demonstrate that a FRED-like envelope naturally emerges whenever the
mass-supply history undergoes a smooth rise followed by a power-law decay,
\textcolor{black}{in physically motivated realizations where such a decay is
present}, while the fine temporal structure arises from shell-to-shell
fluctuations that generate the multi-peaked profiles commonly observed in GRB
prompt emission.

Two cases were explored: 1) In the \emph{mass-driven} scenario, the \textcolor{black}{time-dependent
mass-supply rate} primarily modulates the masses of the ejected shells while the
ejection rate remains nearly constant. Individual pulses tend to maintain
comparable widths while their amplitudes follow the envelope imposed by the
\textcolor{black}{mass-supply} history. 2) In the \emph{rate-driven} scenario, in turn, the shell masses remain roughly constant and the \textcolor{black}{time-dependent mass-supply rate} modulates the
shell widths and consequently the ejection intervals. In this case, pulses
broaden and dim during phases of high \textcolor{black}{mass-supply rate} and
narrow as the \textcolor{black}{mass-supply rate} declines.

Two restrictions of the present implementation should be noted. \textit{First},
the mapping between the \textcolor{black}{time-dependent mass-supply history} and
the engine outflow in both explored cases is still phenomenological. A more
physically grounded connection requires GRMHD simulations of a time-dependent
accretion disk and jetted outflow. \textit{Second}, the temporal evolution of
the spectral peak energy $E_p(t)$ is imposed externally, albeit with physical
motivation. Extensive time-resolved spectral analyses show that $E_p(t)$ tracks
the instantaneous flux, either through a hard-to-soft decay or an
intensity-tracking relation \citep[e.g.,][]{2012ApJ...756..112L,2015A&A...583A.129Y}.

This behavior suggests that the peak energy is closely associated with the
dissipation physics and the engine power. A more complete formulation of
\textcolor{black}{A}MIS should therefore derive the evolution of $E_p$
self-consistently from the \textcolor{black}{engine power, parameterized by the
accretion rate, and the outflow} Lorentz factor, rather than imposing it
externally.

Regarding efficiency, for typical Lorentz-factor fluctuations in the range
$\Gamma \sim \textcolor{black}{10}$--1000, the global radiative efficiency is
$\epsilon_{\rm rad} \sim 0.01$--0.2, consistent with classical internal-shock
expectations, where the dissipated fraction is limited by the Lorentz-factor
contrasts and the distribution of shell masses.

In addition, alternative hybrid scenarios are worth considering. If the outflow
is magnetically dominated, dissipation may proceed through reconnection rather
than shocks. In magnetic reconnection scenarios such as the ICMART model
\citep{2011ApJ...726...90Z}, the \textcolor{black}{time-dependent mass-supply
history} still dictates the overall luminosity, but energy release occurs
through reconnection-driven episodes that are intrinsically more efficient than
internal shocks. Such a picture retains the smooth envelope and rapid
variability characteristic of \textcolor{black}{A}MIS while offering a path toward higher
radiative efficiency.

Future work should aim at connecting \textcolor{black}{time-dependent mass-supply
physics} and jet properties using GRMHD simulations, and at testing the model
quantitatively with large samples of GRBs. Combining the \textcolor{black}{A}MIS
framework with time-resolved fitting of pulse properties, widths, and spectral
evolution may help determine whether \textcolor{black}{mass-supply–modulated
(fallback-motivated where applicable) variability} is a viable explanation for
the prompt phase and what it reveals about the structure of the progenitor and
the central engine. \textcolor{black}{
We emphasize that AMIS describes an intrinsic, engine-level modulation that
provides the primary organizing timescale and luminosity structure of the
prompt emission; jet propagation and viewing-angle effects may introduce
secondary distortions and event-to-event scatter, but do not generically erase
the mass-supply--imprinted temporal trends predicted by the model.
}

\acknowledgments
\textcolor{black}{We are grateful to the referee for their valuable and constructive feedback, which helped improve the manuscript.} R.M thanks S.~Kobayashi for useful discussions and suggestions. The authors used an AI-based tool solely for language polishing. \textcolor{black}{R. Moradi acknowledges support from the Academy of Sciences Beijing Natural Science Foundation (IS24021) and the Institute of High Energy Physics, Chinese(E32984U810).}


\end{document}